\documentclass[preprint,aps,tightenlines,showpacs,nofootinbib,superscriptaddress]{revtex4}
\usepackage{color}
\usepackage{enumerate}
\usepackage{latexsym}
\usepackage[dvips]{graphicx}
\newcommand{\beq}{\begin{eqnarray}}
\newcommand{\eeq}{\end{eqnarray}}
\newcommand{\eq}{eqnarray}

\newcommand{\al}{{\alpha}}
\newcommand{\be}{{\beta}}
\newcommand{\ci}{\cite}

\newcommand{\la}{{\lambda}}
\newcommand{\La}{{\Lambda}}

\newcommand{\om}{{\omega}}

\newcommand{\f}{\frac}
\newcommand{\ra}{\rightarrow}

\begin{document}

\preprint{arXiv:1502.06375v4 [hep-th]}

\title{Black Hole as a Wormhole Factory
}

\author{Sung-Won Kim \footnote{E-mail address: sungwon@ewha.ac.kr}}
\affiliation{Department of Science Education, Ewha Womans University, Seoul, 120-750, Korea}

\author{Mu-In Park \footnote{E-mail address: muinpark@gmail.com, Corresponding author}}
\affiliation{Research Institute for Basic Science, Sogang University, Seoul, 121-742, Korea}

\begin{abstract}
There have been lots of debates about the final fate of an evaporating black hole and the
singularity hidden by an event horizon in quantum gravity. However, on general grounds,
one may argue that a black hole stops radiation at the Planck mass
$(\hbar c/G)^{1/2}\sim 10^{-5} g$, where the radiated energy is comparable to the black hole's
mass. And {also,} it has been argued that there would be a {\it wormhole-like} structure,
known as ``space-time foam", due to large fluctuations below the Planck length
$(\hbar G/c^3)^{1/2}\sim 10^{-33} cm$. In this paper, {as an explicit example},
we {consider}
an exact classical solution which represents nicely those two properties in a recently proposed quantum gravity model based on different scaling dimensions between space and time {coordinates}. The solution, called ``Black Wormhole", consists of two different states{, depending on its mass parameter $M$ and an IR parameter $\om$}: For the black hole state {(with $\om M^2 >1/2$)}, a {\it non-traversable} wormhole occupies the interior region of the black hole around the singularity at the origin, whereas for the wormhole state {(with $\om M^2 <1/2$)}, the interior wormhole is exposed to an outside observer as the {black hole} horizon is disappeared from evaporation. The black hole state becomes thermodynamically stable as it approaches to the merge point where the interior wormhole throat and the black hole horizon merges, and the Hawking temperature vanishes at the exact merge point {(with $\om M^2 =1/2$)}. This solution suggests the ``Generalized Cosmic Censorship" by the existence of a wormhole-like structure which protects the {naked} singularity even after the black hole evaporation. One could understand the would-be wormhole
inside the black hole horizon as the results of microscopic wormholes
created by {``negative"} energy quanta which have entered the
black hole horizon in Hawking radiation processes; {\it the quantum black
hole could be a wormhole factory !} It is found that this
{speculative} picture may be consistent with the recent
``$ER=EPR$" proposal for {resolving} the
black hole entanglement debates.
\end{abstract}


\maketitle

\newpage


It is widely accepted that general relativity (GR) would not be appropriate for describing the small scale structure of space-time. For example, GR, when combined with quantum mechanics, provides a length scale $l_P=(\hbar G/c^3)^{1/2}\sim 10^{-33} cm$, which may provide an absolute limitation for the measurements of space-time distances \ci{Sale}. Actually, this is the length scale on which quantum fluctuations of the space-time are expected to be of order of unity. On the other hand, the singularity theorem, stating the necessary existence of singularities, where the classical concept of space and time breaks down, at certain space-time domains with some reasonable assumptions in GR \ci{Penr}, may be regarded as an indication of the incompleteness of GR.

These circumstances may provide strong motivations to find the quantum theory of gravity which can treat the above mentioned problems of GR. Actually, the necessity of quantizing the gravity has been argued in order to have a consistent interaction with a quantum system \ci{Eppl}. Moreover, it has been also shown that even small quantum gravitational effects dramatically change the characteristic features of a black hole so that it can emit radiations though the causal structures of the classical geometry is unchanged in the semiclassical treatment \ci{Hawk}.

However, as the black hole becomes smaller and smaller by losing its mass from
emitting particles, the semiclassical treatment becomes inaccurate and one can
not ignore the back reactions of the emitted particles on the metric and the
quantum fluctuations on the metric itself anymore. Actually, regarding the back
reaction effects, one can argue that a black hole stops radiations at the Planck
mass $m_P=(\hbar c/G)^{1/2}\sim 10^{-5} g$, where the radiated energy is comparable to the black hole's mass, since a black hole can not radiate more energy than it
has, via the pair creation process near the black hole horizon. This implies
that the black hole should become thermodynamically stable as it becomes smaller
and finally has the vanishing Hawking temperature at the smallest black hole mass.
It seems that this should be one of verifiable predictions that any theory of
quantum gravity make {\ci{Bona}}.
Moreover, according to large fluctuations of metric below the Planck length $l_P$ \ci{Sale}, the {\it wormhole-like} structure, known as ``space-time foam", has been proposed by Wheeler \ci{Whee}. This may be another verifiable prediction of the quantum gravity also.

The purpose of this paper is to {consider}
an exact classical solution{, as an explicit example,} which represents nicely those two properties in a recently proposed quantum gravity model, known  as Ho\v{r}ava gravity, based on different scaling dimensions between space and time {coordinates}. The solution, called ``Black Wormhole", consists of two different states {depending on its mass parameter $M$ and an IR (infrared) parameter $\om$}: For the black hole state {(with $\om M^2 >1/2$)}, a {\it non-traversable} wormhole occupies the interior region of the black hole around the singularity at the origin, whereas for the wormhole state {(with $\om M^2 <1/2$)}, the interior wormhole is exposed to an outside observer as the {black hole} horizon is disappeared from evaporation. The black hole state becomes thermodynamically stable as it approaches to the merge point where the interior wormhole throat and the black hole horizon merges, and the Hawking temperature vanishes at the exact merge point {(with $\om M^2 =1/2$)}.

The solution suggests that, in quantum gravity, the `conventional' cosmic
censorship can be generalized even after black hole evaporation by forming a
wormhole throat around the used-to-be singularity. In GR, black hole and
wormhole are quite distinct objects due to their completely different causal
structures. But the claimed ``Generalized Cosmic Censorship" suggests that
the end state of a black hole is a wormhole, not a naked singularity.
This may correspond to a foam{-like} nature of space-time at short length
scales. {Furthermore, one could understand the would-be wormhole
inside the black hole horizon as the results of microscopic wormholes
created by ``negative" energy quanta which have entered the black hole
horizon in Hawking radiation processes so that {\it the quantum black
hole could be a wormhole factory.} It is found that this {speculative}
picture may be consistent with the recent ``$ER=EPR$" proposal for
{resolving} the
black hole entanglement debates.}


To see how this {picture} can {arise} 
explicitly, we consider the
Ho\v{r}ava gravity
which has been proposed as a four-dimensional, renormalizable, higher-derivative quantum gravity without ghost problems, by adopting different scaling dimensions for space and time {coordinates} in {UV (ultraviolet)} energy regime, $[t]=-1, [{\bf x}]=-z$ with the dynamical critical exponents $z \geq 3$, ``at the expense of Lorentz invariance" \ci{Hora}. We formally define the quantum gravity by a path integral
\begin{\eq}
Z=\int [{\cal D}g_{ij}] [{\cal D} N_i] [{\cal D} N] e^{iS/\hbar}
\end{\eq}
with the proposed action ($z=3$ is considered, for simplicity), up to the surface terms,
\begin{\eq}
S &= & \int dt d^3 x
\sqrt{g}N\left[\frac{2}{\kappa^2}\left(K_{ij}K^{ij}-\lambda
K^2\right)-\frac{\kappa^2}{2 \nu^4}C_{ij}C^{ij}+\frac{\kappa^2
\mu}{2 \nu^2}\epsilon^{ijk} R^{(3)}_{i\ell} \nabla_{j}R^{(3)\ell}{}_k
\right.
\nonumber \\
&&\left. -\frac{\kappa^2\mu^2}{8} R^{(3)}_{ij}
R^{(3)ij}+\frac{\kappa^2 \mu^2}{8(3\lambda-1)}
\left(\frac{4\lambda-1}{4}(R^{(3)})^2-\Lambda_W R^{(3)}+3
\Lambda_W^2\right)+\f{\kappa^2 \mu^2 \om}{8(3\lambda-1)} R^{(3)}\right]\ , \label{horava}
\end{\eq}
and with the ADM decomposition of the metric,
\begin{\eq}
ds^2=-N^2 c^2 dt^2+g_{ij}\left(dx^i+N^i dt\right)\left(dx^j+N^j
dt\right) ,
\end{\eq}
the extrinsic curvature,
\begin{\eq}
 K_{ij}=\frac{1}{2N}\left(\dot{g}_{ij}-\nabla_i
N_j-\nabla_jN_i\right) ,
 \end{\eq}
the Cotton tensor,
\begin{\eq}
 C^{ij}=\epsilon^{ik\ell}\nabla_k
\left(R^{(3)j}{}_\ell-\frac{1}{4}R^{(3)} \delta^j_\ell\right)\ ,
 \end{\eq}
and coupling constants, $\kappa,\lambda,\nu,\mu, \La_W, \om$. The last term in the action (\ref{horava}) represents a ``soft" violation{, with the IR parameter $\om$,} of the ``detailed balance" condition in \ci{Hora} and this modifies the {IR} behaviors so that Newtonian gravity limit exists \ci{Keha,Park,Kiri}.

The proposed action is not the most general form for a power-counting renormalizable gravity, compatible with the assumed foliation preserving {\it Diff} but it is general enough to contain all the known GR solutions, and the qualitative features of the solutions are expected to be similar \ci{Lu,Keha,Park,Kiri}. Here, originally, the {\it non-relativistic} higher-derivative deformation{s were} introduced from the technical reason of the necessity of renormalizable interactions without {the ghost problem} which exists in relativistic higher-derivative theories \ci{Hora}. But {we further remark that, which has not been well emphasized before,} the (UV) Lorentz violation {\it might} have {a} more fundamental reason in our quantum gravity set-up since this may be consistent with the existence of the absolute minimum length $l_P$ which does not depend on the reference frames, {violating the usual {\it relativistic} length contraction.\footnote{It could be also possible that the Lorentz violation occurs only ``dynamically" at some level in quantum gravity. For some extensive discussions about this possibility, see \ci{Will}.}}

\begin{figure}
\includegraphics[width=7.3cm,keepaspectratio]{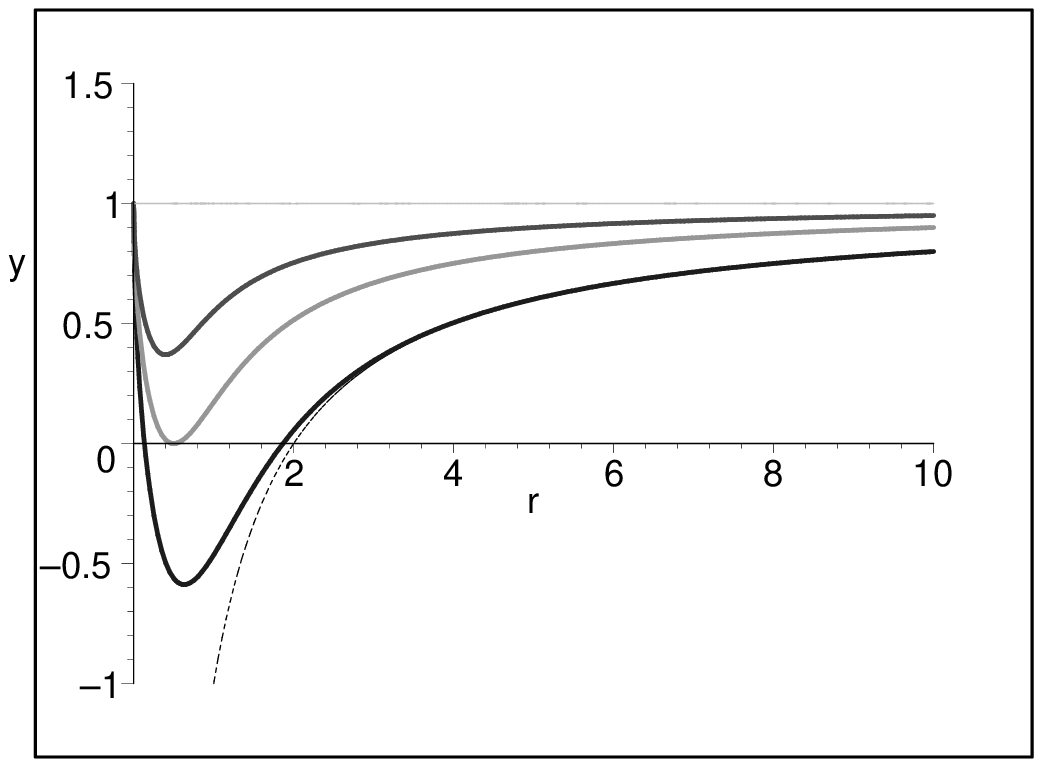} \qquad \qquad
\includegraphics[width=7.3cm,keepaspectratio]{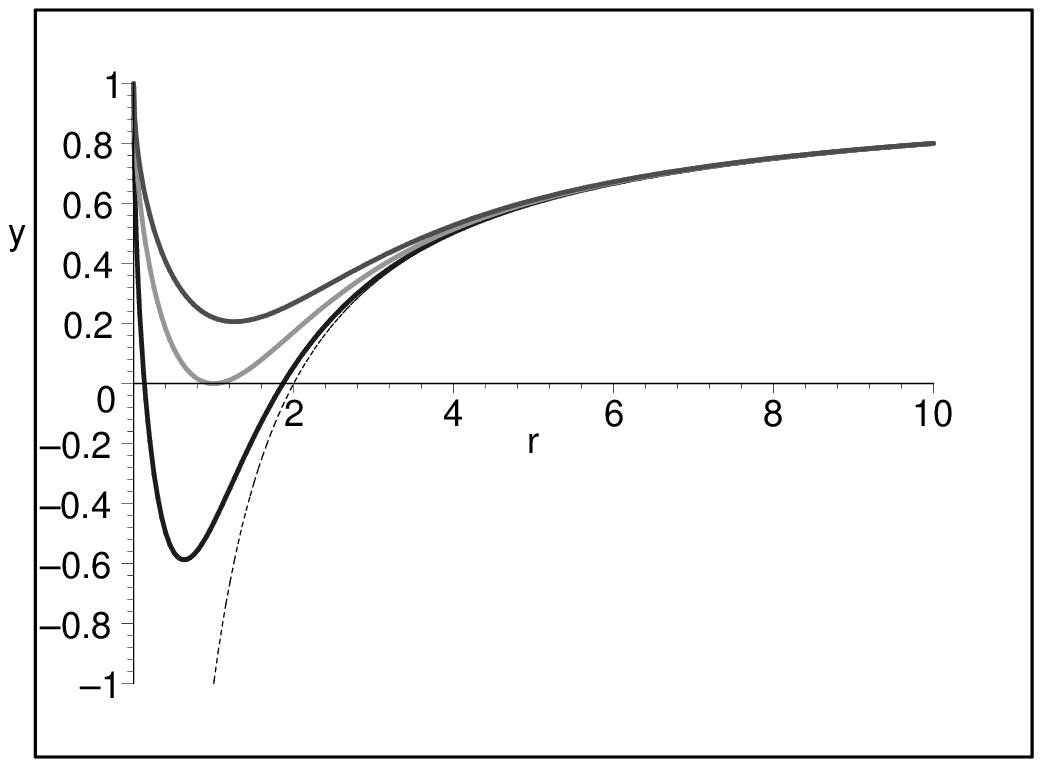}
\caption{Plots of $f(r)$ in Ho\v{r}ava gravity for varying $M$ with a fixed $\om$ (left) and for varying $\om$ with a fixed $M$ (right). In particular, in the left we consider $M=0, 0.25, 0.5, 1$ (top to bottom solid curves) with $\om=2$, and in the right $\om=0.25, 0.5, 2$ (top to bottom solid curves) with $M=1$, in contrast with Schwarzschild solution for $M=1$ in GR (dotted curve).} \label{fig:KS}
\end{figure}
For the simplest case of static, {{\it i.e.},} non-rotating, uncharged black holes, where only the {last} three terms in the action (\ref{horava}) are relevant, the exact solutions have been found completely for arbitrary values of coupling constants, $\la, \La_W$, and $\om$ \ci{Lu,Keha,Park,Kiri}. However, for the present purpose we only consider a simple example of $\la=1, \La_W=0$,
\begin{\eq}
  ds^2=-N(r)^2 c^2 dt^2+\frac{dr^2}{f(r)}+r^2
\left(d\theta^2+\sin^2\theta d\phi^2\right) \label{BH1}
\end{\eq}
with
\begin{\eq}
N^2=f=1+\om r^2-\sqrt{r[\omega^2  r^3 + 4 \om M]}
\label{BH2}
\end{\eq}
so that the standard Einstein-Hilbert action and the asymptotically flat,
Schwarzschild black hole {solution} are recovered in the IR limit, {\it i.e.},
$N^2=f=1 - {2 M}/{r} +{\cal O}(r^{-4})$ with
$c^2={\kappa^4 \mu^2 \omega}/{32},~G={\kappa^2 c^2}/{32 \pi}$ \ci{Keha,Park}. Here $M$ denotes
the `$(G/c^2)\times ADM~ mass$' and the positive {IR} parameter $\om$ controls the strength of
higher-derivative corrections so that the limit $\om \ra \infty, \mu \ra 0$ with `$\mu^2 \om =fixed$' corresponds to GR limit.\footnote{The importance of this limit in a more extensive context will be discussed elsewhere \ci{Argu}.}

One remarkable property of the solution is that there is an inner horizon $r_-$ as well as the outer horizon $r_+$, which solves $f(r_{\pm})=0$, at
\begin{\eq}
r_{\pm} =M \left(1 \pm \sqrt{1-\f{1}{2 \om M^2}} \right)
\end{\eq}
as the result of higher (spatial) derivatives (Fig. 1): The higher derivative terms act like some (non-relativistic) effective matters in the conventional Einstein equation so that there is some ``repulsive" interaction at short {distances}. Moreover, even though the metric converges to the Minkowski's flat space-time at the origin $r=0$, {{\it i.e.}, $N^2=f=1 - 2\sqrt{\om M r} +\om r^2+{\cal O}(r^{7/2})$}, its derivative is not continuous so that there is a curvature singularity at $r=0$, which may be captured by the singularity of $R \sim r^{-3/2},~R^{\mu \nu \al \be} R_{\mu \nu \al \be} \sim r^{-3}$. {So, even} though the singularity at $r=0$ is a time-like line ({\it i.e.}, time-like singularity) and is milder than that of Schwarzschild black hole, $R^{\mu \nu \al \be} R_{\mu \nu \al \be} \sim r^{-6}$ (and also Reissner-Nordstr\"{o}m's $R^{\mu \nu \al \be} R_{\mu \nu \al \be} \sim r^{-4}$) and surrounded by the (two) horizons provided
\begin{\eq}
\om M^2 \geq \f{1}{2},
\end{\eq}
this might indicate that the proposed gravity does not completely resolve the singularity problem of GR still, classically.

This circumstance looks inconsistent with the cosmology solution, where the initial singularity does not exist whence there exist the higher-derivative effects, {\it i.e.}, non-flat universes \ci{Keha,Park}. Moreover, the black hole singularity becomes naked for $\om M^2 <1/2$ so that the cosmic censorship might not work in this edge of solution space{, even if $M$ is positive definite.}

But according to recent BottaCantcheff-Grandi-Sturla (BGS)'s construction of
wormholes, it seems that there is another possibility for resolving the unsatisfactory circumstance \ci{Cant}. What they found was that there exits a wormhole solution also, in addition to the naked singularity solution for $\om M^2 <1/2$, without introducing additional (exotic) matters at the throat. Their obtained wormhole solution
\begin{\eq}
  ds^2=-N_{\pm}(r)^2 c^2 dt^2+\frac{dr^2}{f_{\pm}(r)}+r^2
\left(d\theta^2+\sin^2\theta d\phi^2\right) \label{Wormhole}
\end{\eq}
with
\begin{\eq}
N_{\pm}^2=f_{\pm}=1+\om_{\pm} r^2-\sqrt{r[\om_{\pm}^2  r^3 + 4 \om_{\pm} M_{\pm}]}
\label{Wormhole2}
\end{\eq}
is made of two coordinate patches, each one covering the range $[r_0, +\infty)$ in one universe and the two patches joining at the wormhole throat $r_0$, which is defined as the minimum of the radial coordinate $r$.

{In the conventional approach for (traversable) wormholes, there are
basically two unsatisfactory features \ci{Morr,Viss}. First, we do not know much about
the mechanism for {a} wormhole formation. This is in contrast to the black
hole case, where the gravitational collapse of ordinary matters, like stars, can form
a black hole, if its mass is enough. Second, the usual steps of `constructing' wormholes
are too artificial as summarized by the following three steps: (1). Prepare two or
several universes; (2). Connect the universes by cut and paste of their throats;
(3). Put the needed ``exotic" matters, which violate the energy conditions, to the
throats so that Einstein's equations are satisfied.}

{However, in the new approach, the problem of artificiality of a
wormhole construction is avoided by observing}
that the solution is smoothly joined at the throat, so that the additional compensating (singular or non-singular) matters are not needed, {if} the metric and its derivatives {are} continuous at the throat. But, if we {further} consider the reflection symmetric two universes, {\it i.e.}, $f_+(r)=f_-(r), N^2_+(r)=N^2_-(r)$, with $\om_+=\om_-\equiv \om, M_+=M_- \equiv M$, the only possible way of smooth patching at the throat is \footnote{In BGS's paper \ci{Cant}, it is claimed that the junction condition (\ref{junction}) is not always necessary for $\la=1$ but more general class of wormhole solutions could be possible. But, this is only for the case of singular, $\delta$-function discontinuities in the equations of motions. Whereas, for the non-singular discontinuities at the throat, which can not be properly treated in the BGS's analysis, the condition (\ref{junction}) is still essential for our wormhole construction.}
\begin{\eq}
\left.\f{df_{\pm}}{dr}\right|_{r_0} =0
\label{junction}
\end{\eq}
and the throat radius is obtained as
\begin{\eq}
r_0=\left(\f{M}{2 \om} \right)^{1/3}.
\end{\eq}

{Here, it is important to note} that the throat is located always inside the
black hole horizon, {\it i.e.}, $r_-< r_0< r_+$ for
$\om M^2 >1/2$ so that it is unobservable to an outer observer, whereas the wormhole
throat emerges for $\om M^2 <1/2$, instead of a naked singularity (Fig. 2). For a
fixed $\om$, the throat radius, after emerging from the coincidence with the
extremal black hole radius, $r_0=r_+=r_-=M$, decreases monotonically as $M$ decreases
and finally vanishes for $M=0$, {\it i.e.}, Minkowski vacuum (Fig. 2, left).
{This is the situation that has been assumed in BGS's paper \ci{Cant}.}
But, {on the other hand}, for a fixed $M$ and varying $\om$, one finds
that the throat radius increases again indefinitely as $\om$ decreases. This means
that the wormhole size can be quite large when the coupling constant $\om$, which
could flow under renormalization group, becomes smaller at quantum gravity regime,
like the Planck size black hole or wormhole
(Fig. 2, right) \footnote{ In order that this feature can be seen explicitly, it is important to
consider the correct mass parameter $M$ \ci{Cai}, rather than another form of an
integration constant $\beta=4 \om G M/c^2$ \ci{Lu,Park}}.
In the actual quantum
{gravity} process, like black hole evaporations, $M${, due to Hawking radiations,} as well as $\om${, due to renormalization group flow with the change of energy scales,} can vary so that we need to consider some combinations of situations of the left and right in Fig. 2{, by extending the original interpretation.}
\begin{figure}
\includegraphics[width=7.3cm,keepaspectratio]{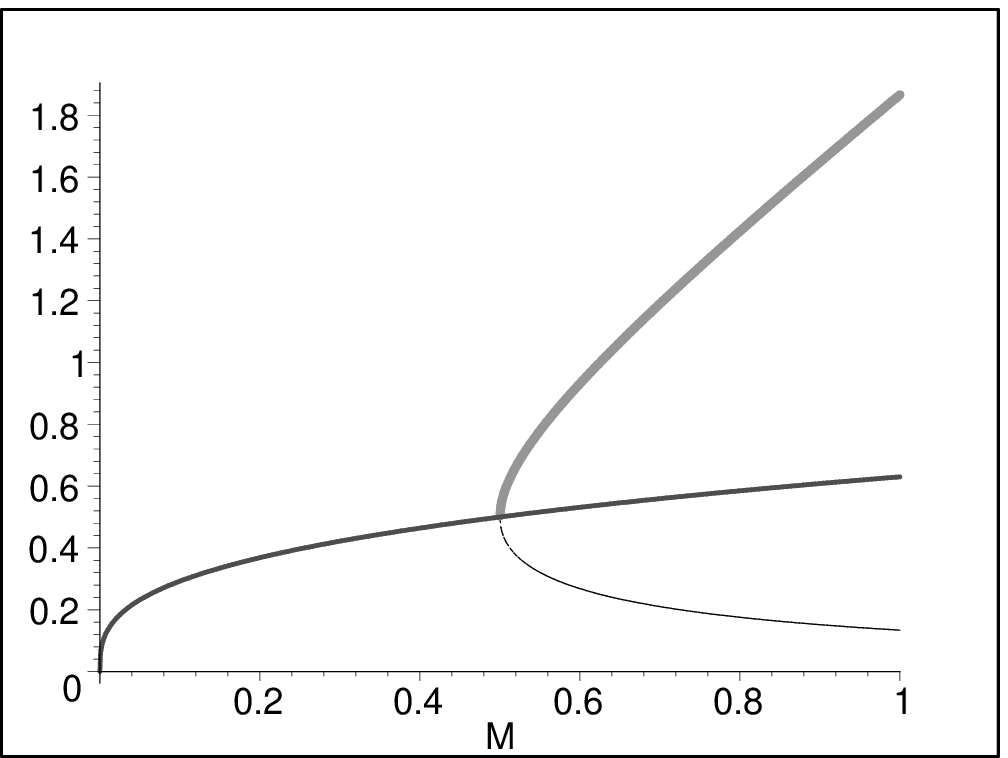} \qquad \qquad
\includegraphics[width=7.3cm,keepaspectratio]{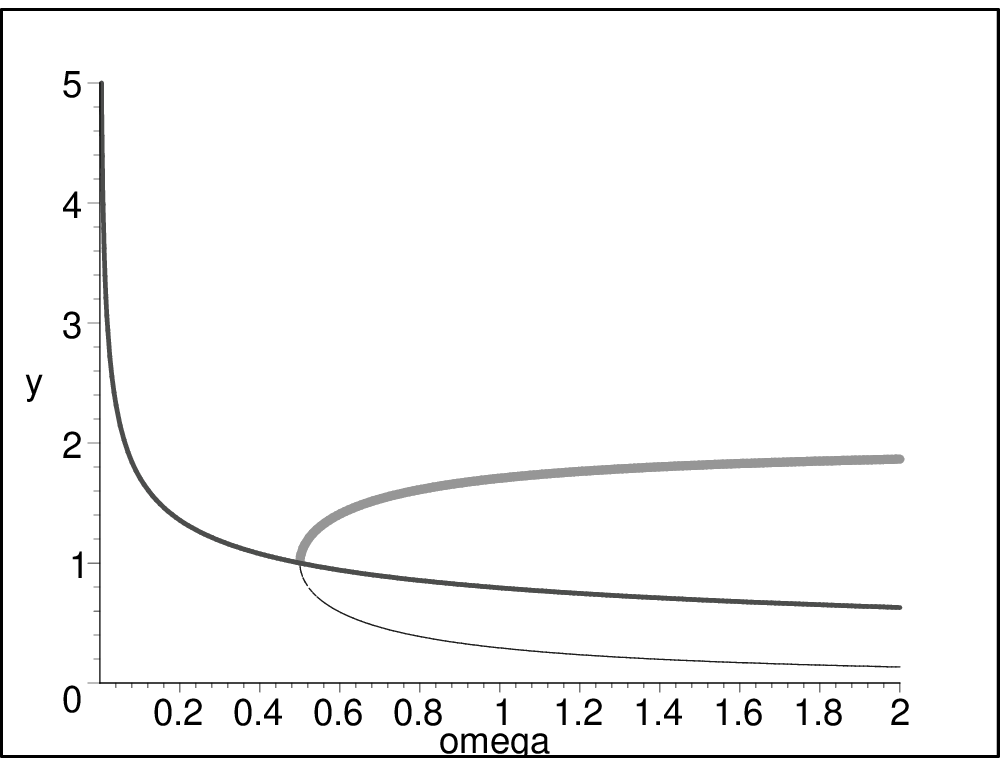}
\caption{Black hole horizons (top ($r_{+}$) and bottom ($r_{-}$) curves) and wormhole throat (middle ($r_0$) curve) radii for varying $M$ with a fixed $\om=2$ (left), and for varying $\om$ with a fixed $M=1$ (right).} \label{fig:throat}
\end{figure}

So,
the wormhole solution is obtained, {without the problem of artificiality,}
when the black hole horizon disappears for $\om M^2 <1/2$.
{Whereas,} 
the solution for $\om M^2 >1/2$ with the throat inside the horizon is {\it not}
the ``traversable" wormhole since the throat will be still inside the
{black hole} horizon of the ``mirror" black hole in another universe. This is
similar to the Einstein-Rosen bridge \ci{Eins} but the difference is that the
throat may not coincide with the black hole horizon generally in
{the present} case. Moreover, in
{the present} case the throat is located at the fixed ``time" $r_0$ so that any time-like trajectories should
{meet}
the throat if exits (Fig. 3). This means that the black hole solution (\ref{Wormhole}),  (\ref{Wormhole2}) with the ``time-like" throat for $\om M^2 >1/2$ should be considered as a physically distinct object from the black hole solution (\ref{BH1}), (\ref{BH2}) having the (hidden) singularity at $r=0$. And in order to avoid a rather strange situation that the wormhole throat ``suddenly" emerges from the extremal black hole which has a singularity {at $r=0$} still, it would be {\it natural} to consider the time-like throat in (\ref{Wormhole}), (\ref{Wormhole2}) for $\om M^2 >1/2$ as the ``would-be" wormhole throat. In order to be distinguished from the usual black hole solution (\ref{BH1}), (\ref{BH2}), we may call {the solution (\ref{Wormhole}), (\ref{Wormhole2})} as the ``Black Wormhole" solution.
\begin{figure}
\includegraphics[width=12cm,keepaspectratio]{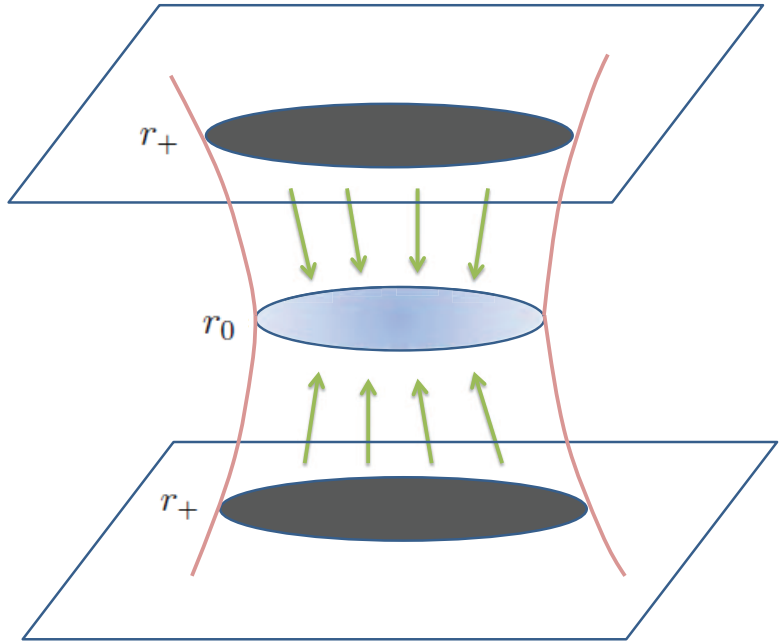}
\caption{A black hole pair which are connected by a wormhole throat $r_0$ inside the black hole horizons $r_+$. The arrows represent the time-like geodesics inside the horizons. } \label{fig:throat}
\end{figure}

Now, we have a completely regular vacuum solution, without the curvature singularities, which interpolates between the black hole state
{for $\om M^2 >1/2$}
and wormhole state {for $\om M^2 <1/2$.}
This seems to support Wheeler's foam picture of the quantum space-time in
quantum gravity but as a real static, not as a virtual time-dependent, wormhole.
And in
our quantum gravity theory, where the concept of horizon emerges {at} low energies, escaping the horizons is not impossible for high energy particles with Lorentz-violating dispersion relations. But for low energy point of view, without probing the interior structure of real black holes, the observable consequences of the black wormhole solution would be expressed in the form of
a ``Generalized Cosmic Censorship"\footnote{In \ci{Cant}, this notion was considered as a ``complementary" between the two mechanisms to censure the singularities.},
suggesting that ``the naked singularity does not appear
{still} by forming wormholes even after the horizons disappear in quantum gravity" though, before evaporation, there would be no naked singularity by the existence of horizons as usual.

Another important {new} implication of the black wormhole solution to low energy observers is that there would be transformations between black holes and wormholes
\footnote{
Hayward has suggested a similar black hole-wormhole transformation on the
grounds of ``trapping" horizons which  may describe black holes and wormholes
unifiedly \ci{Hayw}. But in his framework, assuming the {\it bifurcating}
black hole horizons is essential and it is not clear how to extend his
framework to our case of non-bifurcating horizons of extremal state where
wormhole throat and extremal black hole horizon coincides. Moreover, he
did not suggest that interior structure of a black hole might be changed
after the transformation from a wormhole so that the singularity might
not occur in the black hole state also.}, which is known to be impossible
in GR, due to the no-go theorem for topology change \ci{Gero}. In other
words, once a (primordial) wormhole is formed in the quantum gravity
regime, due to quantum fluctuation in the early universe, it may
evolve into a black hole state by the combinations of Fig. 2 (right)
from renormalization group flows to GR and Fig. 2 (left) from accretion
of matters. It would be interesting to see whether this {could}
be a mechanism for the primordial black holes and supermassive black holes, which are believed to be formed very early in the universe and distinguished from the stellar-mass black holes which are {(or believed to be)} generated from collapsing stars.


On the other hand, {according to the black wormhole solution}, once a
black hole is formed in {our} quantum gravity, it always has the
would-be wormhole inside the horizon but this inside wormhole is exposed to outer
observers when the horizon disappears after the complete evaporation. But we know
that the inside wormhole is absent in the GR limit, as can be seen in Fig. 2 (right).
Then, where does the inside wormhole come from in quantum gravity regime ?
{This is the question about the physical mechanism of a wormhole
formation, which has been lacking.} It seems that the only possible answer to
this question {could}
be found in the Hawking radiation process, which involves virtual pairs of
particles near the event horizon, one of the pair enters into the black hole
while the other escapes: The escaped particle is observed as a real particle
with a {\it positive} energy with respect to an observer at infinity and then,
the absorbed particle must have a {\it negative} {energy} in order that the energy is
conserved \ci{Hawk}. This implies that the negative energy particles that
fall into black hole could be {an ``exotic matter"} source for
a wormhole formation inside the horizon. {But one can easily
discover that this would be quite implausible in GR since negative
masses (or energies) repel ({\it i.e.}, produce the outward accelerations of),
as positive masses attract ({\it i.e.}, produce the inward accelerations of),
all other bodies regardless of their (positive or negative) masses from
the {\it equivalence principle}. We have never observed the negative mass
object yet and it could produce some strange phenomena, known as
``the runaway motion" of a pair of positive and negative mass particles,
but the repulsive nature of gravity for negative masses is another remarkable
consequence of GR \ci{Bond}. In the language of the Newtonian approximation,
where the gravitational potential of a spherically symmetric body is given
by $\varphi(r)=(f(r)-1)/2$, that property is indicated by the fact that
$df/dr$, which is related to the (radial) acceleration $a=-d \varphi/dr=-(1/2)(df/dr)$, is always positive for the potential of a positive mass $M>0$,
whereas $df/dr$ is always negative for the potential of a negative mass
$M<0$. This property implies that negative mass could not form a
(stable) structure, {\it naturally}. This could explain why it would not be
possible to form a wormhole by collapsing of exotic matters, in particular,
negative energy matters, including the negative energy particles that fall
into black hole in GR, in contrast to the case of a black hole formation.
Actually, previously
the negative energy particle has been thought to result in just reducing the
black hole mass by some compensation
process with positive energy sources, which may exist inside the horizon, for the positive black hole mass.}

{However, the situation is quite different in our quantum
gravity context, where the gravity becomes weaker at short distance and changes
its (attractive or repulsive) nature after passing the surface of $df/dr=0$,
which we call the ``zero-gravity surface" for convenience. In other words, for
the positive black hole mass $M>0$, the gravity becomes repulsive inside
the zero-gravity surface (Fig. 1, solid curves). This now implies that even the
positive mass could not form a structure {\it inside the zero-gravity surface},
naturally, in contrast to the outside the zero-gravity surface where the its
gravity is still attractive so that a stable structure can be formed, as in GR.}

\begin{figure}
\includegraphics[width=7.3cm,keepaspectratio]{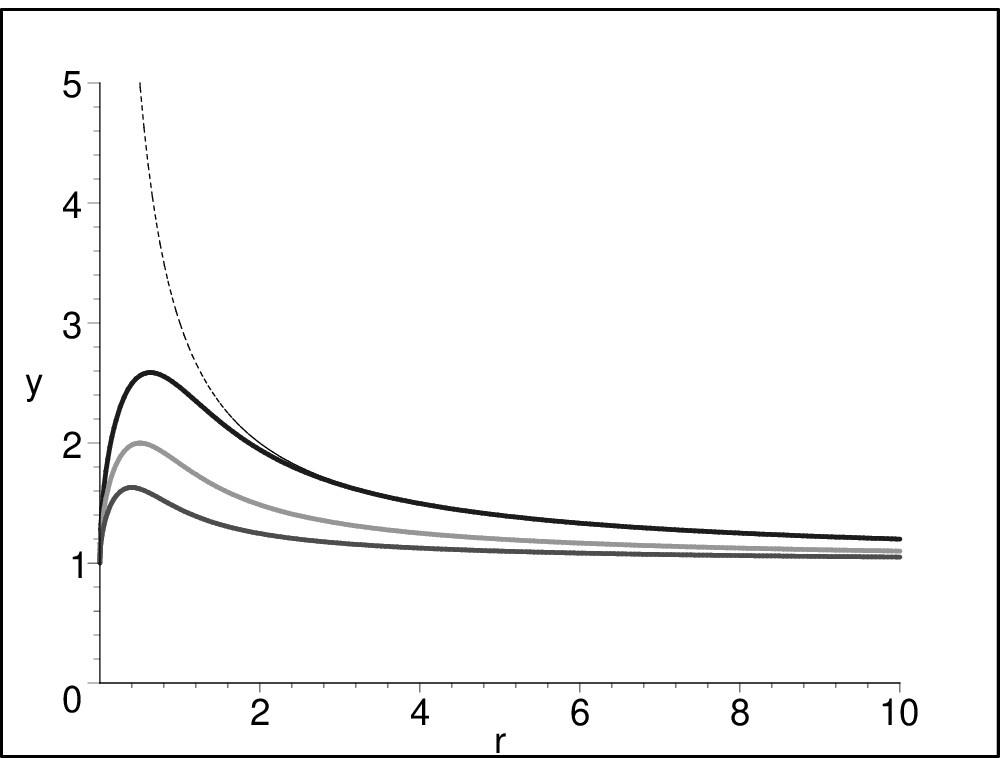}
\qquad \qquad
\includegraphics[width=7.3cm,keepaspectratio]{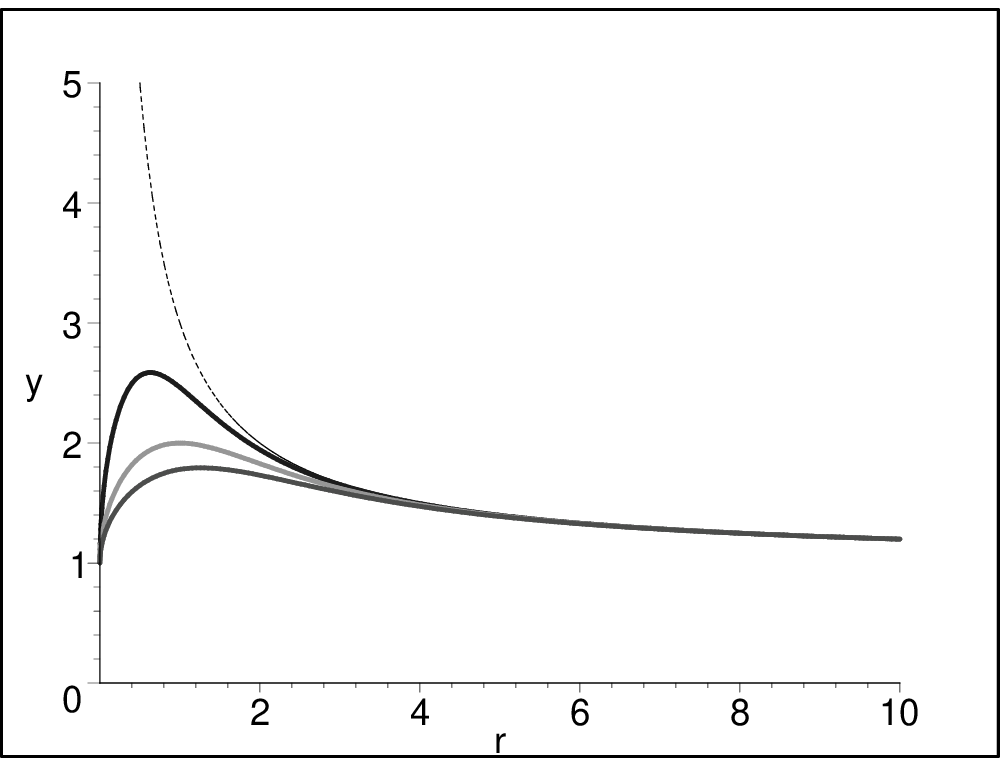}
\caption{Plots of the {\it negative mass}
solution of
$f(r)$ in Ho\v{r}ava gravity for varying $M$ with a fixed $\om$ (left) and for varying $\om$ with a fixed $M$ (right). In particular, in the left we consider $M=-0.25, -0.5, -1$ (bottom to top solid curves) with $\om=-2$, and in the right $\om=-0.25, -0.5, -2$ (bottom to top solid curves) with $M=-1$, in contrast with the negative mass Schwarzschild solution for $M=-1$ in GR (dotted curve).} \label{fig:negative_mass}
\end{figure}
{In order to see how this problem could be resolved dramatically by
the negative masses in our quantum gravity context, we now turn to consider the
{\it negative mass} solution in Ho\v{r}ava gravity, which has never been studied in
the literature. In order to find a negative mass solution, one might first
try to consider $M<0$ for the solution (7) but one can easily find that this could not
be the right solution: The solution is {\it not defined} below a certain radius,
where the quantity inside the square root become zero for $\om>0$ or shows a
different asymptotic ($r\ra \infty$) behavior, {\it i.e.,} not asymptotically
flat, $N^2=f=1+2 \om r^2+2 M/r+{\cal O}(r^{-4})$, for $\om<0$. However, by noting that the solution of the metric ansatz (6) is unique up to the ($\pm$) sign in front of the square root in (7) generally, we find that the negative mass solution as
\begin{\eq}
N^2=f=1+\om r^2 ``+"\sqrt{r[\omega^2  r^3 + 4 \om M]}
\label{Negative_M}
\end{\eq}
for the different sign of the square root term, compared to the solution (7).
This solution recovers the Schwarzschild solution in the IR limit for
$\om<0$, as (7) does for $M>0$ and $\om>0$ (Fig. 4); for $\om>0$,
the solution (14) neither recovers the Schwarzschild solution nor
the metric is defined for the whole space region. [But in this case, in order that GR is recovered, we need to consider $\mu^2<0$ as in asymptotically de Sitter space \ci{Park} and we will discuss about this again later]

One can prove that the new solution (14) does not have horizons, as can be easily observed in Fig. 4 also but it has a curvature singularity at $r=0$, where the metric is finite but discontinuous, with the same degrees of divergences as the solution (7). So, this solution itself represents a naked singularity as the negative mass Schwarzschild solution does in GR. Actually, this may be one of main obstacle for considering the negative mass solution as a viable solution in GR and this is not much improved for the solution (14) alone, in our Ho\v{r}ava gravity context.

But, as in the positive mass case, we have another (exact) solution of a wormhole, for the reflection symmetric two universes, with the same formula (13) for the throat $r_0$. In other words, in Ho\v{r}ava gravity, there exists a regular, {\it i.e., singularity free}, negative mass solution so that it could be viable but only in the form of a wormhole geometry. But here we will not consider about the negative mass wormhole solution further since its formation mechanism is unclear at present.

Rather, we consider the negative mass solution (14) in order to see how does the negative masses, like the negative energy particles that fall into black hole could interact and form a structure, like the (black) wormhole with a positive mass. Actually, the solution (14) indicates that the gravity of a negative mass is repulsive at large distance but becomes weaker, {\it i.e.}, less-repulsive, at short distance and moreover becomes attractive inside the zero-gravity
surface of $df/dr=0$. This implies that the negative masses could form a structure
naturally, at short distance inside the zero-gravity surface, in contrast to the case of
positive masses which
could not form a structure inside their {\it own} zero-gravity surface as we have explained already.
This is the remarkable consequence of Ho\v{r}ava gravity which could justify
our picture that {\it wormholes are created and sustained by the continuous
inflow of negative energy particles via the Hawking radiation process}, which is
impossible in GR. This is the main claim of this paper and in a more compact
form, this can be expressed as
:{\it The quantum black hole could be a wormhole factory !} }

Interestingly, this may provide a physical origin of the Einstein-Rosen bridge in the recent ``$ER=EPR$" proposal for resolving the issue of entanglement in a black hole spacetime \ci{Mald} claiming that ``the black hole and its Hawking radiation are entangled via Einstein-Rosen bridges" (Fig. 5). Following their proposal, one could understand the would-be wormholes inside the black hole horizon as the results of microscopic wormholes created by negative energy quanta which have entered the black hole horizon but entangled with its Hawking radiated, positive energy, partner quanta, which now lives in a single universe, widely separated from {its} mother black hole. This may be consistent with our reflection symmetric configuration, but it {is} not clear how to extend this picture to more general configurations which can not be interpreted as a configuration in a single universe. Moreover, it is not clear either how the throats of Einstein-Rosen's bridges evolve into our would-be wormhole in detail.
\begin{figure}
\includegraphics[width=12cm,keepaspectratio]{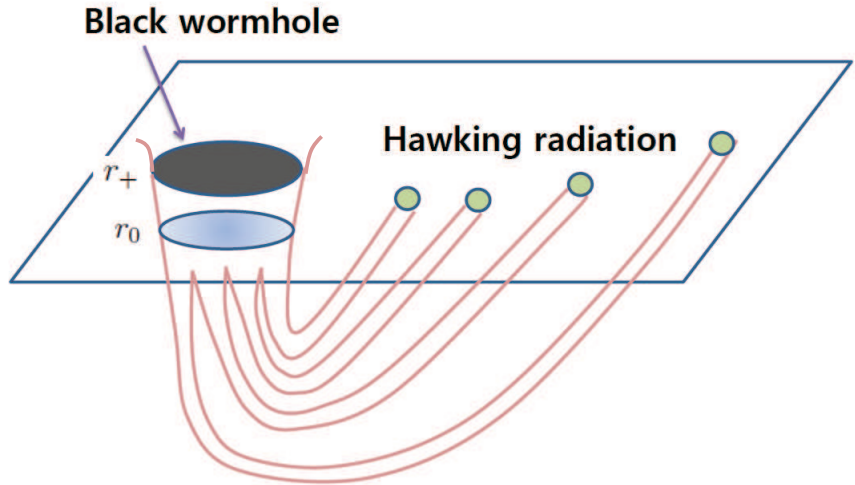}
\caption{The microscopic would-be wormholes created by negative energy quanta which have entered the black hole horizon with its Hawking radiated, positive energy, partner quanta outside the horizon. {This describes the quantum black hole as a factory of microscopic wormholes which would merge to a macroscopic black wormhole solution, in conformity with "ER=EPR" proposal.}} \label{fig:throat}
\end{figure}


In conclusion, we have {considered}
a vacuum and static {\it black wormhole} solution which
is regular, {\it i.e.}, singularity-free, and interpolates between the black hole state
{for $\om M^2 >1/2$}
and wormhole state
{for $\om M^2 <1/2$}
through the coincidence state of
an extremal black hole and a {(kind of)} Einstein-Rosen bridge {for $\om M^2 =1/2$}. From this, we have suggested the transformation between the black hole and wormhole states, and its resulting generalized cosmic censorship. {And furthermore, we have argued that the would-be wormholes inside the black hole horizon could be understood as the results of microscopic wormholes created by ``negative" energy quanta which have entered the black hole horizon in Hawking radiation processes so that {\it the quantum black hole could be a wormhole factory.}}

But then, ``Can the transformation really occur dynamically ?" In order to
answer to this question, we first consider the transformation from a black
hole state to a wormhole state. Of course,
{there is} some obstacle in GR {context} since
the black hole should
{pass} \footnote{Recently, there have been quite active researches and debates about whether a non-extremal black hole can ``jump over" the extremality, by capturing a particle with appropriate parameters \ci{Hube}. In particular, a quantum jump or tunneling seems to be a quite promising mechanism for breaking extremal black holes in our quantum gravity set-up \ci{Mats}.} the extremal black hole to become a wormhole
state {but the
extremal black hole has vanishing Hawking radiation and temperature so
that it is believed as the stable ground state in the black hole states.}
{However recently, certain classical instability, known as {\it Aretakis instability}, for extremal black holes have been found in GR so that the usual belief may not be quite correct \ci{Aret}. A heuristic argument about this instability suggests that the inner horizon instability of near-extremal black holes \ci{Pois} might cause an instability of the coincided inner and outer horizons \ci{Maro}. If this is the case, we may expect a similar instability in our extremal black holes also, due to the (expected) inner horizon instability.\footnote{The inner horizon instability \ci{Pois} may be related to the {\it negative} temperature for the inner horizon, {\it i.e.}, $T_-\equiv (\hbar/4 \pi)(df/dr)_{r_-}<0$ \ci{Park:2006}, which seems to be a quite generic property of the inner horizons, including those of the black holes in Ho\v{r}ava gravity \ci{Park}. }}

Now then, let us consider the transformation from a wormhole state to a
black hole state. In this case, there does not seem to exist any
{classical
obstacle against this transformation, which resembles the collapse of {stars} made of ordinary matter to  make a black hole, where the entropy would increase in the process.} Actually, in the conventional wormhole context with exotic matters in GR, the collapse of the Morris-Thorne wormhole {into a black hole} was observed from non-linear instability under ordinary matter perturbations \ci{Shin}. We expect the similar non-linear instability, though linearly stable, exits for our wormhole state {too} so that it transforms to the extremal black hole state, a seed of large black holes.

{Finally, two further remarks are in order.

First, even though we {have} obtained the solution in a particular quantum gravity model
which is power-counting renormalizable without ghost problems, the features of the
small scale structure seems to be quite generic if {\it the vanishing Hawking
temperature for a Planck mass black hole}, which implying the existence of the wormhole
throat $r_0$ satisfying $df/dr|_{r_0}=0$,
is considered as a verifiable prediction that any theory of quantum gravity makes. For example, so called, ``renormalization group improved black hole spacetimes" would have the similar wormhole structure and our discussion may not be limited to Ho\v{r}ava gravity \ci{Bona}. But we can easily check that it is not applicable to Kerr nor Reissner-Nordstr\"{o}m black holes in GR: In these cases, there are more metric functions or additional matter fields but there is no solution for the throat where all the metric functions or fields join smoothly\footnote{Similarly, it looks like that the existence of the throat for more general black holes with charges or rotations in the Ho\v{r}ava gravity would be very difficult, unless some accidental coincidences occur. This seems to implies that the wormhole throat in the spherically symmetric configurations could be easily destroyed by adding other hairs. Conversely, the wormhole throat could be formed only after losing all the hairs except the mass.}.

Second, we have found that, in order to describe the negative mass solution
(14), we need to consider the coupling constant of $\mu^2<0$ so that one
can recover the well-defined GR parameters $c^2, G>0$ in the IR limit. This does not affect the couplings of the second-derivative terms in the action (2) (note that we are considering the case of $\La_W=0$) but only affects other
UV couplings \ci{Lu}, which differ from those for the positive mass
solution (7). This indicates that the (weak) {\it equivalence principle} may
be violated by UV effects
generally, depending on the sign of particle's mass.
It would be a challenging problem that the runaway motions of a pair of positive and negative mass particles could be avoided from the inequivalent motions of the positive and negative mass particles in UV.}


\section*{Acknowledgments}

This work 
was supported by Basic Science
Research Program through the National Research
Foundation of Korea (NRF) funded by the Ministry
of Education, Science and Technology (2010-
0013054) (SWK), (2-2013-4569-001-1) (MIP).

\newcommand{\J}[4]{#1 {\bf #2} #3 (#4)}
\newcommand{\andJ}[3]{{\bf #1} (#2) #3}
\newcommand{\AP}{Ann. Phys. (N.Y.)}
\newcommand{\MPL}{Mod. Phys. Lett.}
\newcommand{\NP}{Nucl. Phys.}
\newcommand{\PL}{Phys. Lett.}
\newcommand{\PR}{Phys. Rev. D}
\newcommand{\PRL}{Phys. Rev. Lett.}
\newcommand{\PTP}{Prog. Theor. Phys.}
\newcommand{\hep}[1]{ hep-th/{#1}}
\newcommand{\hepp}[1]{ hep-ph/{#1}}
\newcommand{\hepg}[1]{ gr-qc/{#1}}
\newcommand{\bi}{ \bibitem}

\end{document}